\documentclass[aps,pra,twocolumn,amsmath,amssymb,nofootinbib,superscriptaddress]{revtex4-1}
\usepackage{times}
\usepackage[pdftex]{graphicx}
\usepackage{dcolumn}
\usepackage{bm}
\usepackage{amsmath}
\usepackage{indentfirst}
\usepackage{float}
\usepackage[colorlinks]{hyperref}
\usepackage[dvipsnames]{xcolor}

\usepackage{makecell}
\usepackage[normalem]{ulem}

\newcommand{\gammaeff}{\gamma_{{\rm eff}}}
\newcommand{\Dop}{\mathcal{D}}
\newcommand{\Hop}{\hat{H}}
\newcommand{\Oop}{\hat{O}}
\newcommand{\LDHop}{\hat{H}_{{\rm LD}}}
\newcommand{\rhoop}{\hat{\rho}}
\newcommand{\rhoss}{\hat{\rho}_{{\rm st}}}
\newcommand{\bop}{\hat{a}}
\newcommand{\bdop}{\hat{a}^{\dagger}}
\newcommand{\xop}{\hat{x}}
\newcommand{\nop}{\hat{n}}

\newcommand{\nss}{\bar{n}_{{\rm st}}}
\newcommand{\hc}{{\rm H.c.}}
\newcommand{\trace}{{\rm tr}}
\newcommand{\im}{{\rm i}}

\newcommand{\const}{{\rm const }}
\newcommand{\gcc}[1]{{\color{black}{#1}}}

\begin{document}

\title{Steady state phonon occupation of EIT cooling: higher order calculations}

\author{Shuo Zhang}
\affiliation{Henan Key Laboratory of Quantum Information and Cryptography, Zhengzhou,
Henan 450000, China}

\author{Tian-Ci Tian}
% \email{119254691@qq.com}
\affiliation{Henan Key Laboratory of Quantum Information and Cryptography, Zhengzhou,
Henan 450000, China}

\author{Zheng-Yang Wu}
% \email{chengyangair@outlook.com}
\affiliation{Henan Key Laboratory of Quantum Information and Cryptography, Zhengzhou,
Henan 450000, China}

\author{Zong-Sheng Zhang}
% \email{562971598@qq.com}
\affiliation{Henan Key Laboratory of Quantum Information and Cryptography, Zhengzhou,
Henan 450000, China}

\author{Xin-Hai Wang}
% \email{1066559569@qq.com}
\affiliation{Henan Key Laboratory of Quantum Information and Cryptography, Zhengzhou,
Henan 450000, China}

\author{Wei Wu}
\affiliation{Department of Physics, College of Liberal Arts and Sciences, National University of Defense Technology, Changsha 410073, China
Interdisciplinary Center for Quantum Information, National University of Defense Technology, Changsha 410073, China}

\author{Wan-Su Bao}
\email{bws@qiclab.cn}
\affiliation{Henan Key Laboratory of Quantum Information and Cryptography, Zhengzhou,
Henan 450000, China}

\author{Chu Guo}
\email{guochu604b@gmail.com}
\affiliation{Henan Key Laboratory of Quantum Information and Cryptography, Zhengzhou,
Henan 450000, China}

\begin{abstract}
Electromagnetically induced transparency (EIT) cooling has established itself as one of the most widely used cooling schemes for trapped ions during the past twenty years. Compared to its alternatives, EIT cooling possesses important advantages such as a tunable effective linewidth, a very low steady state phonon occupation, and applicability for multiple ions. However, existing analytic expression for the steady state phonon occupation of EIT cooling is limited to the zeroth order of the Lamb-Dicke parameter. Here we extend such calculations and present the explicit expression to the second order of the Lamb-Dicke parameter. We discuss several implications of our refined formula and are able to resolve certain \gcc{difficulties} in existing results.
\end{abstract}

\date{\today}
\maketitle

\address{}

\vspace{8mm}

\section{Introduction}
Laser cooling of a trapped ion into its motional ground state is a key
step towards various applications in quantum simulation~\cite{PorrasCirac2004b,PorrasCirac2004a,LeibfriedWineland2003,BermudezPlenio2013,RuizCampo2014,RammHartmut2014,GuoPoletti2015,GuoPoletti2016,GuoPoletti2017,GuoPoletti2017b,GuoPoletti2018,XuPoletti2019,PanDavidson2020,WuChen2019}, quantum computing~\cite{CiracZoller1995,LanyonRoos2011,KielpinskiWineland2002}, \gcc{quantum metrology~\cite{ChouRosenband2010,LudlowSchmidt2015,HuntemannPeik2016}, as well as testing foundations of quantum mechanics~\cite{Wineland2013}}.
% \cite{prl-74-4091,science-319-1808,prl-104-070802,apr-6-021314}.
Till now there exists a variety of cooling schemes that work in the Lamb-Dicke (LD) regime in which the Lamb-Dicke parameter $\eta$ satisfies $\eta \ll 1$. Those schemes could be roughly categorized into sideband cooling~\cite{MonroeWineland1995,RoosBlatt1999}, dark-state cooling~\cite{MorigiKeitel2000,EversKeitel2004,RetzkerPlenio2007,CerrilloPlenio2010,AlbrechtPlenio2011,ZhangChen2012,YiYang2013,ZhangChen2014,LuGuo2015,YiGu2017,CerrilloPlenio2018} and feedback cooling~\cite{SteixnerZoller2005,RablZoller2005,BushevZoller2006,ZhangBao2017}.

Up to date, EIT cooling is perhaps one of the most widely used dark-state cooling scheme in trapped ions experiments~\cite{RoosBlatt2000,LinWineland2013,KampschulteMorigi2014,LechnerRoos2016,ScharnhorstSchmidt2018,JordanBollinger2019,FengMonroe2020,QiaoKim2020,HuangLan2021}. Compared to its alternatives, EIT cooling possesses several outstanding advantages. First, compared to the traditional sideband cooling, the resolved sideband condition, namely the linewidth $\gamma$ of the excited state should be much smaller than the \gcc{axial
secular} trap frequency $\nu$, is not required in EIT cooling. Second, compared to other dark-state cooling schemes as well as feedback cooling schemes, EIT cooling, in its ideal realization, only requires three internal energy levels and two static lasers, making it an experimentally friendly scheme. Lastly, the internal ground state is prepared as a dark-state of the carrier transition, eliminating one of the two major heating mechanisms (the other one is the blue sideband transition), and therefore a very low steady state phonon occupation could be reached. Moreover, since the internal excitation of the ion is transparent to the lasers, in a multiple ion setup, EIT cooling could easily cool down one phonon mode without heating up the others. Due to the same reason, it is recently pointed out that EIT cooling could actually be implemented in the strong sideband coupling regime, speeding up the cooling rate by more than one order of magnitude without significant increase in the steady state phonon occupation~\cite{ZhangGuo2021,LiChen2021}.

In this work we are primarily interested in the steady state phonon occupation defined as $\nss = \trace(\nop \rhoss)$, where $\nop$ is the phonon number operator and $\rhoss$ is the \gcc{density operator of the system at} steady state. In literatures, $\nss$ is often expressed as an $\eta$-independent term plus a contribution of the order $\eta^2$, that is, $\nss = \const + O(\eta^2)$. Here we note that, the recoil energy due to one photon emission is of the order $\eta^{2}$. As an example, for sideband cooling $\nss^{{\rm sb}}=\left(\alpha+\frac{1}{4}\right)\left(\frac{\gamma}{2\nu}\right)^{2} + O\left(\eta^{2}\right)$, where the geometry factor $\alpha = 2 / 5 $ for dipole transition, while for standing wave sideband cooling $\nss^{{\rm swsb}}=\frac{1}{4}\left(\frac{\gamma}{2\nu}\right)^{2} + O\left(\eta^{2}\right)$~\cite{CiracZoller1992}. In comparison, $\nss^{{\rm eit}}$ for EIT cooling is first given in Ref.~\cite{MorigiKeitel2000} as
\begin{align}\label{eq:eitnwsc}
\nss^{{\rm eit}}=\frac{\gamma^{2}}{16\Delta^{2}}+O\left(\eta^{2}\right),
\end{align}
where $\Delta$ is the detuning for both lasers used in EIT cooling. Eq.(\ref{eq:eitnwsc}) also reflects the flexibility of EIT cooling since $\Delta$ is highly tunable, while in the case of sideband cooling, the \gcc{axial
secular} trap frequency $\nu$ is often a given constant.  Subsequent dark-state cooling schemes utilize more complicated energy level structures and laser setups to suppress both the carrier and blue sideband transitions, thus completely eliminating the $\eta$-independent term on the right hand side of Eq.(\ref{eq:eitnwsc}) and reach a steady state phonon occupation of the order $O\left(\eta^{2}\right)$~\cite{EversKeitel2004,CerrilloPlenio2010,AlbrechtPlenio2011,ZhangChen2012,CerrilloPlenio2018}, which means that the ion could be cooled down to recoil or even subrecoil temperature.

To this end we point out that the expression in Eq.(\ref{eq:eitnwsc}) poses at least two theoretical \gcc{difficulties}: 1) In the EIT cooling setup, the excited state $\vert e\rangle$ of linewidth $\gamma$ dissipates to both the two ground states $\vert g\rangle$ and $\vert r\rangle$, with dissipation rates $\gamma_g$ and $\gamma_r$ satisfying $\gamma = \gamma_g + \gamma_r$, as can be seen from Fig.~\ref{fig:fig1}(a). However the dark state is a certain combination of $\vert g\rangle$ and $\vert r\rangle$, and cooling is essentially induced by an effective dissipation rate $\gammaeff$ from $\vert e\rangle$ to this dark state. In certain parameter regime (which does not violates the EIT cooling conditions) one could even have $\gammaeff \approx 0$ (This statement will become clear later in the main text), and cooling should not be possible. Such effect can not be predicted from Eq.(\ref{eq:eitnwsc}) since it only depends on $\gamma$; 2) $\Delta$ is usually highly tunable in experiments. Therefore it is possible to make it very large such that $\frac{\gamma^2}{16\Delta^2}$ is comparable or even less than the $O(\eta^2)$ term, in which case the recoil temperature could already be achieved without eliminating the blue sideband. However this statement can only be made precise once we have the exact expression for the $O(\eta^2)$ term. Therefore in this work we perform more refined calculation of $\nss^{{\rm eit}}$ which gives us an explicit expression to the order of $\eta^2$, with which we are able to resolve these two \gcc{difficulties}.

This paper is organized as follows. In Sec.~\ref{sec:model} we derive the master equation for EIT cooling in a slightly different representation compared to the previous works, which would be more convenient for the derivations in this work. In Sec.~\ref{sec:results}, we solve the master equation to obtain our main result, namely the analytic expression for the $O(\eta^2)$ term in Eq.(\ref{eq:eitnwsc}). We verify our result by comparing it with the exact numerical solutions of the master equation. We also discuss several implications of it and use it to resolve the two above mentioned \gcc{difficulties} of Eq.(\ref{eq:eitnwsc}). We conclude in Sec.~\ref{sec:summary}. Since in this work we only focus on EIT cooling, we will eliminate the superscript in the expression of $\nss^{{\rm eit}}$ in the following.

\section{Master equation for EIT cooling}\label{sec:model}

\begin{figure}
\includegraphics[width=\columnwidth]{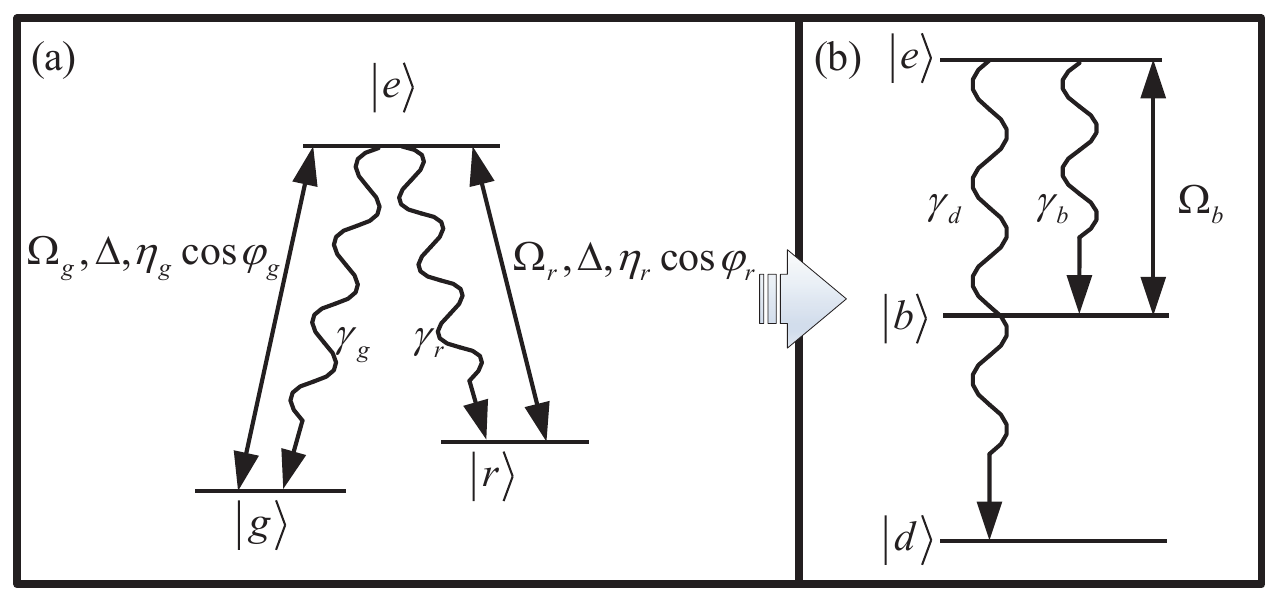}
\caption{(a) The standard three-level configuration of EIT cooling. The system consists of
a dissipative excited state $\left|e\right\rangle $ and two ground
states $\left|g\right\rangle $ and $\left|r\right\rangle $. The transitions
$\left|e\right\rangle \leftrightarrow\left|g\right\rangle $ and $\left|e\right\rangle \leftrightarrow\left|r\right\rangle $
are induced by two external lasers. (b) The dynamics of internal
degrees of freedom can be understood in the basis of $\left\{ \left|e\right\rangle ,\left|b\right\rangle ,\left|d\right\rangle \right\} $, where the dark
state $\vert d\rangle$ decouples from $\left|b\right\rangle $ and $\left|e\right\rangle $, and the bright state $\left|b\right\rangle $ is coupled to $\left|e\right\rangle $ with an effective coupling strength $\Omega_{b}$. }
\label{fig:fig1}
\end{figure}

In standard EIT cooling setup, an ion of mass $M$ is confined in
a harmonic trap with \gcc{axial
secular} trap frequency $\nu$. As shown in Fig.~\ref{fig:fig1}(a), the excited state $\vert e\rangle$ is coupled to two ground states $\vert g\rangle$ and $\vert r\rangle$ by two lasers with frequencies $\omega_{g}$ and $\omega_{r}$, and with Rabi frequencies $\Omega_g$ and $\Omega_r$. The angles of the lasers with respect to the motional axis are $\varphi_g$ and $\varphi_r$ respectively. The excited state $\vert e\rangle$ dissipates to $\vert g\rangle$ and $\vert r\rangle$ with rates $\gamma_g$ and $\gamma_r$. The energy differences between $\vert e\rangle$ and $\vert g\rangle$, between $\vert e\rangle$ and $\vert r\rangle$ are denoted as $\Delta_{eg}$, $\Delta_{er}$ respectively.

The dynamics of the system is described by the Lindblad master equation~\cite{GoriniSudarshan1976,Lindblad1976} (We take $\hbar = 1$ throughout this work)
\begin{align}\label{eq:mastereq}
\frac{\textrm{d}}{\textrm{d}t}\rhoop(t)=-\im\left[\Hop,\rhoop(t)\right]+\Dop(\rhoop(t)).
\end{align}
Here $\rhoop(t)$ is the density operator of the system at time $t$. The Hamiltonian $\Hop$ takes the form
\begin{align}
\Hop = & \nu \bdop\bop-\Delta\left|e\right\rangle \left\langle e\right|\nonumber \\
 & +\left(\frac{\Omega_{g}}{2}\left|e\right\rangle \left\langle g\right|e^{\im k_{g}\xop\cos(\varphi_{g})}+\hc\right)\nonumber \\
 & +\left(\frac{\Omega_{r}}{2}\left|e\right\rangle \left\langle r\right|e^{\im k_{r}\xop\cos(\varphi_{r})}+\hc\right),
\end{align}
with $\bdop$ ($\bop$) being the creation (annihilation) operator of the ion's motional
state (phonon), $\xop = \frac{1}{\sqrt{2M\nu}}(\bdop + \bop)$ being the position operator, and the wave numbers $k_{g}=\omega_g/c,k_{r}=\omega_r/c$ ($c$ is the speed of light).
% and $\varphi_{g}$ $\left(\varphi_{r}\right)$ being the angle between
% the motional axis and cooling (coupling) laser.
As one of the EIT cooling conditions, the detunings of both lasers are tuned to the same, that is,
\begin{align}\label{eq:detune}
\Delta=\omega_{g}-\Delta_{eg}=\omega_{r}-\Delta_{er}.
\end{align}
The LD parameters $\eta_g$ and $\eta_r$ related to the two lasers are defined as
% Expanding $\Hop$ to the first order of the LD parameters $\eta_g$ and $\eta_r$ defined as
\begin{align}
\eta_{g}=k_{g}\sqrt{\frac{1}{2M\nu}},\eta_{r}=k_{r}\sqrt{\frac{1}{2M\nu}}.
\end{align}
Expanding $\Hop$ to the first order of $\eta_g$ and $\eta_r$, we get
\begin{align}\label{eq:LDH}
\LDHop = & \nu \bdop\bop-\Delta\left|e\right\rangle \left\langle e\right|\nonumber \\
 & +\left(\frac{\Omega_{g}}{2}\left|e\right\rangle \left\langle g\right|+\frac{\Omega_{r}}{2}\left|e\right\rangle \left\langle r\right|+\hc\right)\nonumber \\
 & +\left[\im\eta_{g}\cos(\varphi_{g})\frac{\Omega_{g}}{2}\left|e\right\rangle \left\langle g\right|\left(\bdop+\bop\right)+\hc\right]\nonumber \\
 & +\left[\im\eta_{r}\cos(\varphi_{r})\frac{\Omega_{r}}{2}\left|e\right\rangle \left\langle r\right|\left(\bdop+\bop\right)+\hc\right].
\end{align}
% \begin{align}\label{eq:LDH}
% \LDHop = & \nu \bdop\bop-\Delta\left|e\right\rangle \left\langle e\right|\nonumber \\
%  & +\left(\frac{\Omega_{g}}{2}\left|e\right\rangle \left\langle g\right|+\frac{\Omega_{r}}{2}\left|e\right\rangle \left\langle r\right|+\hc\right)\nonumber \\
%  & +\sum_{j=g,r} \left[\im\eta_{j}\cos\varphi_{j}\frac{\Omega_{j}}{2}\left|e\right\rangle \left\langle j\right|\left(\bdop+\bop\right)+\hc\right].
% \end{align}
The dissipative part $\Dop$ will only be kept to the zeroth order of the LD parameters since the next non-vanishing term would contribute only to the $4$-th order of the LD parameters~\cite{CiracZoller1992}, namely we take $\Dop \approx \Dop_{0}$ with
\begin{align}\label{eq:LDD}
\Dop_{0}(\rhoop)=\sum_{j=g,r}\frac{\gamma_{j}}{2}\left(2\left|j\right\rangle \left\langle e\right|\rhoop\left|e\right\rangle \left\langle j\right|- \{\rhoop, \left|e\right\rangle \left\langle e\right|\} \right).
\end{align}
We note that $\Dop_0$ denotes the usual spontaneous emission when the atomic motion is neglected.

\begin{figure}
\includegraphics[width=0.9\columnwidth]{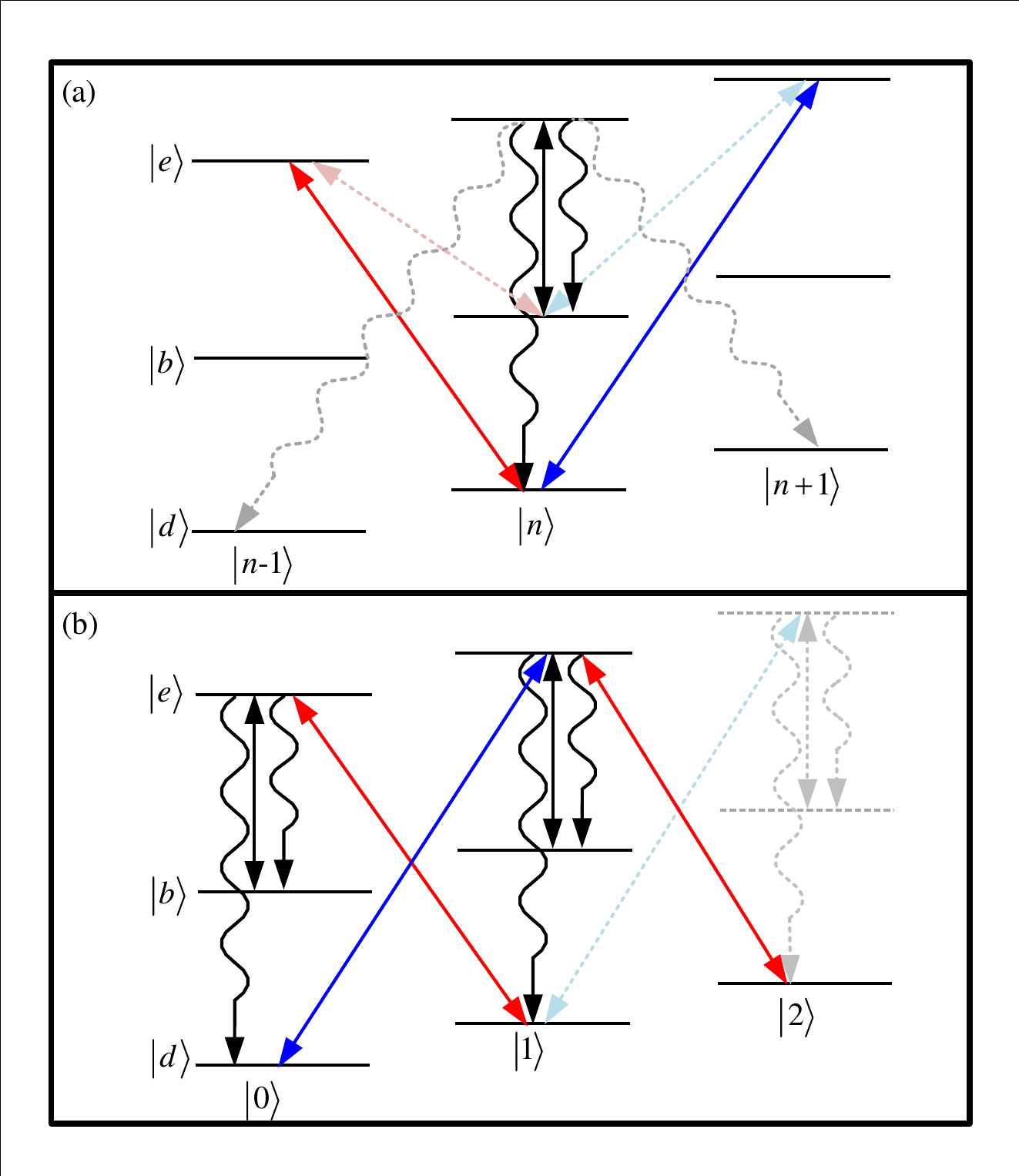}
\caption{The EIT cooling dynamics in the $\{\vert d\rangle, \vert b\rangle, \vert e\rangle \}$ representation. (a) All the possible transitions up to $\eta^2$ are shown in \gcc{solid lines}, which include $\left|d,n\right\rangle \leftrightarrow\left|e,n\pm1\right\rangle $,
$\left|b,n\right\rangle \leftrightarrow\left|e,n\right\rangle $,
and dissipations $\left|e,n\right\rangle \rightarrow\left|b,n\right\rangle $,
$\left|e,n\right\rangle \rightarrow\left|d,n\right\rangle $. Meanwhile,
the contributions of the sideband transitions $\left|b,n\right\rangle \leftrightarrow\left|e,n\pm1\right\rangle $
as well as the effective dissipations $\left|e,n\right\rangle \rightarrow\left|d,n\pm1\right\rangle $
are of the order $O\left(\eta^{4}\right)$, and are shown in \gcc{dashed lines}.
(b) In the steady state of EIT cooling, the population is mostly in the state $\left|d,0\right\rangle $. We thus limit ourself in the subspace $\mathcal{V}_{0}$ spanned by the \gcc{$7$ energy levels most closely related to this state} as shown by the \gcc{solid lines}.}
\label{fig:fig2}
\end{figure}

Under the condition in Eq.(\ref{eq:detune}), we can perform a unitary transformation on the subspace $\{\vert g\rangle, \vert r\rangle\}$ spanned by states $\vert g\rangle$ and $\vert r\rangle$ as
% When the detunings of the two laser are the same, the steady state
% of the internal d.o.f is in the dark state
\begin{align}
\left|d\right\rangle &=\cos(\theta)\left|g\right\rangle -\sin(\theta)\left|r\right\rangle; \label{eq:darkstate} \\
\left|b\right\rangle &=\sin(\theta)\left|g\right\rangle +\cos(\theta)\left|r\right\rangle, \label{eq:brightstate}
\end{align}
with $\tan(\theta)= \Omega_g / \Omega_{r}$. Then $\vert d\rangle$ is a dark state in that it decouples from the subspace spanned by $\vert b\rangle$ and $\vert e\rangle$ if only the internal degrees of freedom are considered. $\vert b\rangle$ is referred to as the bright state since it is coupled to $\vert e\rangle$. In the usual derivation of EIT cooling, $\vert b\rangle$ and $\vert e\rangle$ are further diagonalized into two states $\vert \pm\rangle$ such that the internal degrees of freedom are fully diagonalized. The second condition for EIT cooling is chosen as
\begin{align}\label{eq:cond2}
\Delta=\frac{\Omega_{g}^{2}+\Omega_{r}^{2}}{4\text{\ensuremath{\nu}}},
\end{align}
such that the red sideband transition $\vert d, n+1\rangle \leftrightarrow \vert +, n\rangle$ is resonant~\cite{MorigiKeitel2000,Morigi2003}. In the $\{\vert d\rangle, \vert\pm\rangle\}$ representation, the state $\vert -\rangle$ is neglected since it is far off resonant, leaving the two states $\{\vert d\rangle, \vert+\rangle \}$ together with the phonon states closely resembling the standing wave sideband cooling~\cite{ZhangGuo2021}, with an effective dissipation rate from $\vert +\rangle$ to $\vert d\rangle$ denoted as $\gammaeff$. In general $\gammaeff< \gamma$. This physical picture, although nice for a rough understanding since it reduces the model to the well understood standing wave sideband cooling, has one important difficulty that if we directly apply the standing wave sideband cooling formalism, the obtained $\nss$ will only depend on $\gammaeff$ instead of $\gamma$. The truth is that the dynamics between $\vert d\rangle$ and $\vert -\rangle$ can not really be neglected since the internal dynamics in the subspace $\{\vert b\rangle, \vert e\rangle\}$ could still be much faster than that of the red sideband. To correctly obtain a refined expression for $\nss$, one has to take all the internal states into account. We thus work in the $\{\vert d\rangle, \vert b\rangle , \vert e\rangle\}$ representation of the internal states in the following, which is shown in Fig.~\ref{fig:fig2}(a). In this representation we can rewrite $\LDHop$ in Eq.(\ref{eq:LDH}) as
\begin{align}\label{eq:LDH2}
\LDHop = & \nu \bdop\bop-\Delta\left|e\right\rangle \left\langle e\right|+\left(\frac{\Omega_{b}}{2}\left|e\right\rangle \left\langle b\right|+\hc\right)\nonumber \\
 & +\left(\im\eta\frac{\Omega_{d}}{2}\left|e\right\rangle \left\langle d\right|\left(\bop+\bdop\right)+\hc\right),
\end{align}
where $\Omega_{d}=\frac{\Omega_{g}\Omega_{r}}{\sqrt{\Omega_{g}^{2}+\Omega_{r}^{2}}}$, $\Omega_{b}=\sqrt{\Omega_{g}^{2}+\Omega_{r}^{2}}$
and $\eta=\eta_{g}\cos(\varphi_{g})-\eta_{r}\cos(\varphi_{r})$. Here we have neglected the sideband transitions between $\vert b, n\rangle$ and $\vert e, n\pm 1\rangle$ since their contributions are of the order $\eta^4$. We can also rewrite $\Dop_0$ in Eq.(\ref{eq:LDD}) as
\begin{align}\label{eq:LDD2}
\Dop_{0}(\rhoop) = \sum_{j=d, b} \frac{\gamma_{j}}{2}\left(2\left|j\right\rangle \left\langle e\right|\rhoop\left|e\right\rangle \left\langle j\right|- \{\rhoop, \left|e\right\rangle \left\langle e\right|\}\right),
\end{align}
with $\gamma_{d}=\gamma_{g}\cos^{2}(\theta)+\gamma_{r}\sin^{2}(\theta),\gamma_{b}=\gamma_{r}\cos^{2}(\theta)+\gamma_{g}\sin^{2}(\theta)$.

\section{Results and discussions}\label{sec:results}
In the following we compute the steady state solution of Eq.(\ref{eq:mastereq}) to the second order of $\eta$. Since the population is mostly in the state $\left|d,0\right\rangle $
when reaching the steady state of EIT cooling, we restrict Eq.(\ref{eq:mastereq}) to the subspace
$\mathcal{V}_{0}$ spanned by the $7$ states
\begin{align}
\mathcal{V}_{0} = \left\{ \left|d,0\right\rangle ,\left|b,0\right\rangle ,\left|e,0\right\rangle ,\left|d,1\right\rangle ,\left|b,1\right\rangle ,\left|e,1\right\rangle ,\left|d,2\right\rangle \right\},
\end{align}
as shown in Fig.\ref{fig:fig2}(b). \gcc{The reason for this choice is that the states in $\mathcal{V}_{0}$ can be reached by at most one blue sideband transition starting from $\left|d,0\right\rangle$, which is assumed to be much weaker than both the dissipation and the red sideband.} Then the steady state solution $\rhoss$ to Eq.(\ref{eq:mastereq}) in the subspace $\mathcal{V}_0$ can be found by solving the following set of equations
\begin{align}\label{eq:rhoss}
\trace\left(\Oop\mathcal{P}\left(-\im\left[\LDHop,\rhoss\right]+\Dop_{0}(\rhoss)\right)\right) = 0,
\end{align}
where $\mathcal{P}$ is the projection operator into the subspace $\mathcal{V}_{0}$, and $\Oop$ is an operator belonging to the group $\{\left|j\right\rangle \left\langle j\right|, \left|j\right\rangle \left\langle k\right|+\left|k\right\rangle \left\langle j\right|, \im\left|j\right\rangle \left\langle k\right|-\im\left|k\right\rangle \left\langle j\right|\}$ ($j \neq k$), where $\vert j\rangle$ and $\vert k\rangle$ are any of the states in $\mathcal{V}_0$.

% $\left(\left|j\right\rangle ,\left|k\right\rangle =\right.$$\left|d,0\right\rangle ,$
% $\left|b,0\right\rangle ,$ $\left|e,0\right\rangle ,$$\left|d,1\right\rangle ,$$\left|b,1\right\rangle ,$$\left|e,1\right\rangle ,$$\left|d,2\right\rangle ,$$\left.j\neq k\right)$.

Eq.(\ref{eq:rhoss}) contains $49$ equations in total, which is in general not easy to solve directly. However, under the conditions that $\gamma_{g},\gamma_{r}\ll\Delta$, $\eta\Omega_{d}\ll\Omega_{b}$
and Eq.(\ref{eq:cond2}), we find that certain subblock of equations decouples from the rest, thus simplifying the calculations, and we can get the solutions for all the diagonal terms as
\begin{subequations}\label{eq:solutions}
\begin{align}
\rho_{b0,b0} = & \frac{\eta^{2}\Omega_{d}^{2}\left(\gamma_{d}+\gamma_{b}\right)}{4\Omega_{b}^{2}\gamma_{d}}\rho_{d0,d0}; \\
\rho_{e0,e0} = & 0; \\
\rho_{d1,d1} = & \frac{\eta^{2}\Omega_{d}^{2}\left(\gamma_{d}+\gamma_{b}\right)}{4\Omega_{b}^{2}\gamma_{d}}\left(1+\frac{4\gamma_{d}\nu^{2}\left(\gamma_{d}+\gamma_{b}\right)}{\eta^{2}\Omega_{d}^{2}\Omega_{b}^{2}}\right)\rho_{d0,d0}; \\
\rho_{b1,b1} = & \frac{\eta^{2}\Omega_{d}^{2}\left(\gamma_{d}+\gamma_{b}\right)}{4\Omega_{b}^{2}\gamma_{d}}\rho_{d0,d0}; \\
\rho_{e1,e1} = & 0; \\
\rho_{d2,d2} = & \frac{\eta^{2}\Omega_{d}^{2}}{4\Omega_{b}^{2}}\left(\frac{\gamma_{b}}{\gamma_{d}}\right)\rho_{d0,d0}.
\end{align}
\end{subequations}
Here, we have used $\rho_{b0,b0}=\trace\left(\left|b,0\right\rangle \left\langle b,0\right|\mathcal{P}\rho\right)$ and similar for others. The detailed derivation of Eqs.(\ref{eq:solutions}) is in Appendix.~\ref{app:app1}. Using $\rho_{d0,d0}\approx1$, we obtain $\nss$ as
% \begin{align}
% \nss \approx & \rho_{d1,d1}+\rho_{b1,b1}+\rho_{e1,e1}+2\rho_{d2,d2}\nonumber \\
%   \approx & \frac{\gamma^{2}}{4\Delta^{2}}+\frac{\eta^{2}\Omega_{d}^{2}}{2\Omega_{b}^{2}}+\frac{\eta^{2}\Omega_{d}^{2}}{\Omega_{b}^{2}}\left(\frac{\gamma_{b}}{\gamma_{d}}\right)\label{eq:16}
% \end{align}
% \begin{align}\label{eq:nss}
% \nss \approx & \rho_{d1,d1}+\rho_{b1,b1}+\rho_{e1,e1}+2\rho_{d2,d2}\nonumber \\
%   \approx & \frac{\gamma^{2}}{4\Delta^{2}}+\frac{\eta^{2}\Omega_{d}^{2}}{\Omega_{b}^{2}}\left(\frac{1}{2} +\frac{\gamma_{b}}{\gamma_{d}}\right).
% \end{align}
\begin{align}\label{eq:nss}
\nss \approx & \rho_{d1,d1}+\rho_{b1,b1}+\rho_{e1,e1}+2\rho_{d2,d2}\nonumber \\
  \approx & \frac{\gamma^{2}}{16\Delta^{2}}+\frac{\eta^{2}\Omega_{d}^{2}}{\Omega_{b}^{2}}\left(\frac{1}{2} +\frac{\gamma_{b}}{\gamma_{d}}\right).
\end{align}
Eq.(\ref{eq:nss}) is the main result of this work. Compared with Eq.(\ref{eq:eitnwsc}), we can see that it gives the explicit expression to the order of $\eta^2$. In the following we discuss two special parameter regimes which are frequently considered.

First, $\Omega_{g}\ll\Omega_{r}$, and thus $\theta \approx 0 $. In this regime we have $\vert d\rangle \approx \vert g\rangle$, $\vert b\rangle \approx \vert r\rangle$, $\Omega_d \approx \Omega_g$, $\Omega_b \approx \Omega_r$, $\gamma_d \approx \gamma_g$ and $\gamma_b \approx \gamma_r$. Consequently, we have
% \begin{align}\label{eq:nssa}
% \nss &\approx\frac{\gamma^{2}}{4\Delta^{2}}+\frac{\eta^{2}\Omega_{g}^{2}}{\Omega_{r}^{2}}\left(\frac{1}{2}+\frac{\gamma_{r}}{\gamma_{g}}\right) \nonumber \\
% &\approx \frac{4\gamma^{2} \nu^2 }{\Omega_r^{4}}+\frac{\eta^{2}\Omega_{g}^{2}}{\Omega_{r}^{2}}\left(\frac{1}{2}+\frac{\gamma_{r}}{\gamma_{g}}\right) .
% \end{align}
\begin{align}\label{eq:nssa}
\nss &\approx\frac{\gamma^{2}}{16\Delta^{2}}+\frac{\eta^{2}\Omega_{g}^{2}}{\Omega_{r}^{2}}\left(\frac{1}{2}+\frac{\gamma_{r}}{\gamma_{g}}\right).
\end{align}
In usual experimental setup, the condition $\gamma_g \geq \gamma_r$ can be satisfied (if this is not satisfied, one could simply swap the role of $\vert g\rangle$ and $\vert r\rangle$). In this case the coefficient of the $\eta^2$ term on right hand side of Eq.(\ref{eq:nssa}) can often be smaller compared to the first term, and then Eq.(\ref{eq:eitnwsc}) will agree well with the exact solution of Eq.(\ref{eq:mastereq}). Now we note that in Eq.(~\ref{eq:nssa}), the $\eta^2$ term is explicitly dependent on $\gamma_g$ and $\gamma_r$. From a pure theoretical point of view, Eq.(\ref{eq:nssa}) shows that $\nss$ diverges when $\gamma_g \rightarrow 0$, which coincides with intuition that in this limit ground state cooling can not be achieved. This is an effect which can not be predicted from Eq.(\ref{eq:eitnwsc}).
% This is intuitive since in this case the cooling mostly induced by the dissipation channel from $\vert e, n\rangle$ to $\vert g, n\rangle$, and in the limit $\gamma_g \rightarrow 0$, cooling should not be possible.

Second, $\Omega_{g}=\Omega_{r}$ and thus $\theta=\pi /4$. In this case $\left|d\right\rangle $ and $\left|b\right\rangle $ are equal superposition of states $\left|g\right\rangle $ and $\left|r\right\rangle $ as $\left|d\right\rangle =\frac{1}{\sqrt{2}}\left(\left|g\right\rangle +\left|r\right\rangle \right),\left|b\right\rangle =\frac{1}{\sqrt{2}}\left(\left|g\right\rangle -\left|r\right\rangle \right)$, and $\Omega_{d}=\frac{\Omega_{b}}{2}=\frac{1}{\sqrt{2}}\Omega_{r},\gamma_{d}=\gamma_{b}=\frac{\gamma}{2}$. Thus we have
\begin{align}\label{eq:nssb}
\nss \approx\frac{\gamma^{2}}{16\Delta^{2}}+\frac{3}{8}\eta^{2}.
\end{align}
From Eqs.(\ref{eq:nssa}, \ref{eq:nssb}) we can see that for EIT cooling, the contribution from the $\eta^2$ term is not necessarily smaller than from the $\eta$-independent term, since we can tune $\Delta$ to be much larger than $\gamma$ with the EIT cooling conditions in Eqs.(\ref{eq:detune}, \ref{eq:cond2}) still being satisfied. The necessity of using more elaborated dark-state cooling schemes with more lasers is thus obscure since one could already reach the recoil temperature with carefully tuned parameters in standard EIT cooling. To this end, the two \gcc{difficulties} from Eq.(\ref{eq:eitnwsc}) have been resolved based on our analytical result.

\begin{figure}
\includegraphics[width=\columnwidth]{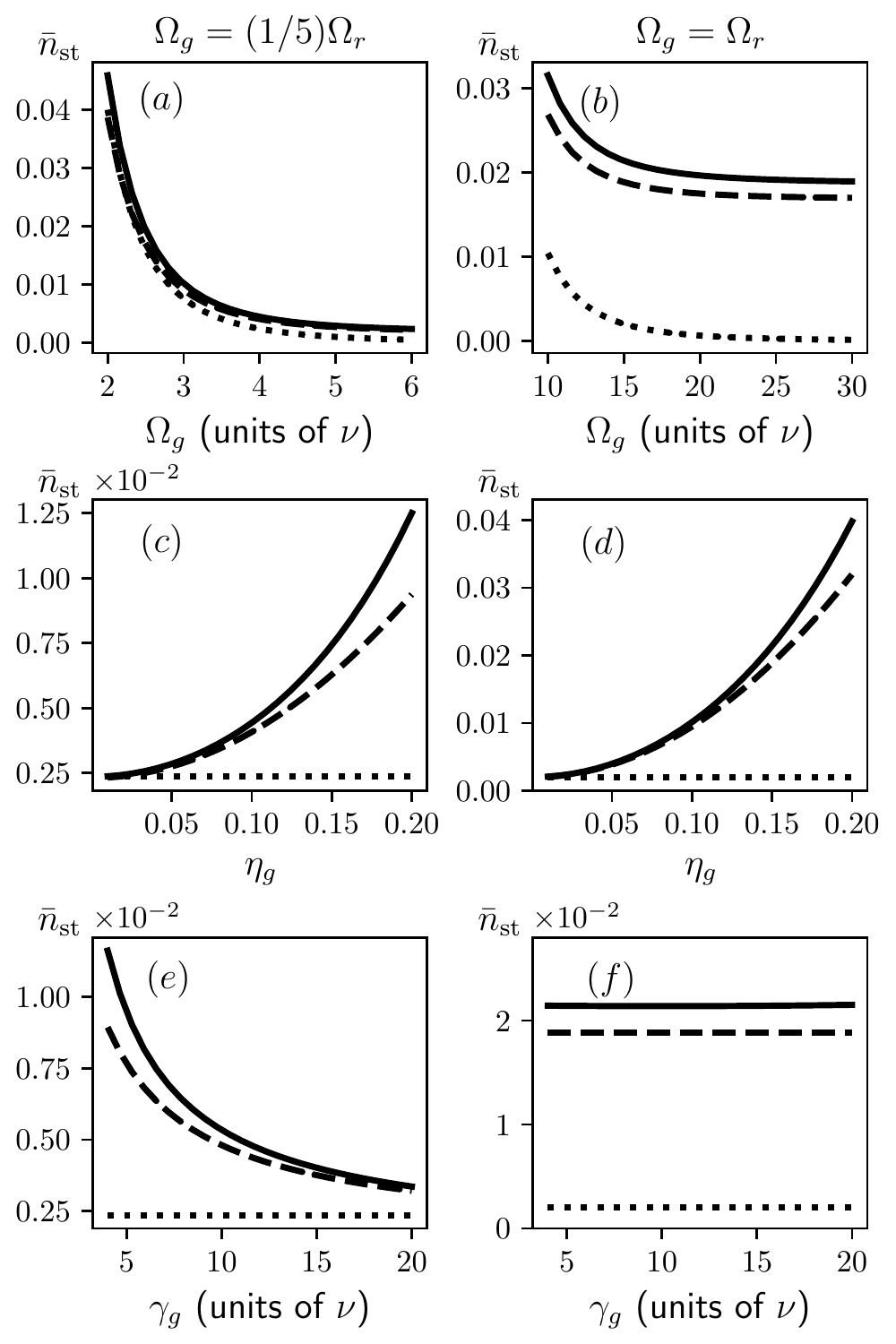}
\caption{Comparison of predicted steady state phonon occupation $\nss$. The solid, dashed, dotted lines in all the panels stand for the exact numerical solutions from Eq.(\ref{eq:mastereq}), predictions from our analytical expression in Eq.(\ref{eq:nss}), and predictions from the previous results in Eq.(\ref{eq:eitnwsc}) with the $O(\eta^2)$ term neglected. (a, c, e) plot $\nss$ as a function of $\Omega_g$, $\eta_g$ and $\gamma_g$ respectively, under the condition $\Omega_g = \Omega_r/5$. (b, d, f) plot $\nss$ as a function of $\Omega_g$, $\eta_g$ and $\gamma_g$ respectively, under the condition $\Omega_g = \Omega_r$. In (c, e) we have used $\Omega_g=4\nu$ while in (d, f) we have used $\Omega_g=15\nu$. In (e, f) we have fixed $\gamma_g + \gamma_r = 20\nu $. The other parameters used (if not specified in the figures) are $\gamma_{g}/\nu=20/3$,
$\gamma_{r}/\nu=40/3$, $\eta_{g}=\eta_{r}=0.15$, $\varphi_{g}=\pi / 4$,
$\varphi_{r}=3\pi /4$. $\Delta$ is chosen according to Eq.(\ref{eq:cond2}). 
% correspond to
% (yellow solid line), and $\rho_{d1,d1}$ (blue solid line), $\rho_{b1,b1}$
% (red solid line), $\rho_{d2,d2}$ (black solid line) as functions
% of $\Omega_{g}$. The other parameters are $\gamma_{g}=40/3\nu$,
% $\gamma_{g}=20/3\nu$, $\eta_{g}=\eta_{r}=0.15$, $\varphi_{g}=\frac{\pi}{4}$,
% $\varphi_{r}=\frac{3\pi}{4}$, (a) $\Omega_{g}=\frac{1}{5}\Omega_{r}$,
% (b) $\Omega_{g}=\Omega_{r}$, and $\Delta$ is chosen to be $\Delta=\frac{\Omega_{b}^{2}}{4\text{\ensuremath{\nu}}}$.
% The dashed lines are corresponding analytical predictions.
}
\label{fig:fig3}
\end{figure}

To verify Eq.(\ref{eq:nss}), we compare it with the exact numerical solution of Eq.(\ref{eq:mastereq}) which is shown in Fig.~\ref{fig:fig3}. To clearly see the effect of the $\eta^2$ term in Eq.(\ref{eq:nss}), we also plot in Fig.~\ref{fig:fig3} the predictions from Eq.(\ref{eq:eitnwsc}) without the $\eta^2$ correction term. More concretely, we perform two sets of simulations, one with $\Omega_g = \Omega_r / 5$ (panels $a, c, e$) and the other with $\Omega_g = \Omega_r$ (panels $b, d, f$). In each set of simulations, we show $\nss$ as a function of $\Omega_g$ (panels $a, b$), $\eta_g$ (panels $c, d$) and $\gamma_g$ (panels $e, f$) respectively. We can see that in all the panels predictions from our analytical expression in Eq.(\ref{eq:nss}) agree better with the exact numerical solutions. Particularly, from panels $(b,d,f)$ we can see that the effect of the $\eta^2$ term is more significant when $\Omega_g=\Omega_r$, this is because in this case the $\Omega_d/\Omega_r$ term in the coefficient of $\eta^2$ reaches its maximum value $1/2$. From panels $(c, d)$ we can see that the $\eta^2$ correction term is more significant when $\eta_g$ becomes larger, in fact it can be several times larger than the contribution of the zeroth term in Eq.(\ref{eq:eitnwsc}) when $\eta_g$ approaches a moderate value of $0.2$. Panel $e$ also reveals the failure of Eq.(\ref{eq:eitnwsc}) when $\Omega_g \ll \Omega_r$ and $\gamma_g$ is small, which is the case we considered in Eq.(\ref{eq:nssa}). Panel $f$ demonstrates the case of Eq.(\ref{eq:nssb}), where our prediction differs from Eq.(\ref{eq:eitnwsc}) by a constant shift $3\eta^2/8$, and we can see that this shift could be much larger than the zeroth order term $\frac{\gamma^2}{16\Delta^2}$.

% The parameters used in our numerical simulations are chosen from a realistic experimental implementation based on the $^{40}$Ca$^{+}$ ion, where the EIT cooling can be implemented on the \gcc{$^{2}S_{1/2}\rightarrow{}^{2}P_{1/2}$} transition. 

% \gcc{In our numerical simulations we have taken the parameters used in $^{40}$Ca$^{+}$ ion experiment as a reference~\cite{RoosBlatt2000,Lechner2016Roos,ScharnhorstSchmidt2018}. }
\gcc{The parameter regimes considered in our numerical simulations could readily be explored in current ion cooling experiments. Taking the $^{40}$Ca$^{+}$ ion as an example, EIT cooling can be implemented on the \gcc{$^{2}S_{1/2}\rightarrow{}^{2}P_{1/2}$} transition.}
Both \gcc{$^{2}S_{1/2}$} and \gcc{$^{2}P_{1/2}$} have two Zeeman sublevels, which constitute
a four-level system. Then one can either choose $\left|e\right\rangle =\left|P,+\right\rangle ,\left|g\right\rangle =\left|S,+\right\rangle ,\left|r\right\rangle =\left|S,-\right\rangle $
or $\left|e\right\rangle =\left|P,-\right\rangle ,\left|g\right\rangle =\left|S,-\right\rangle ,\left|r\right\rangle =\left|S,+\right\rangle $
for EIT cooling. For both choices we have $\gamma_{g}/\gamma_{r}=1/2$, \gcc{which is the case we considered in Fig.\ref{fig:fig3}(a,b,c,d)}. \gcc{The Rabi frequencies $\Omega_g$ and $\Omega_r$ vary case by case, however from Eq.(\ref{eq:nss}) we can see that what really matters for $\nss$ is the ratio $\Omega_g / \Omega_r$, which is often widely tunable in experiments. The Lamb-Dicke parameter could be tuned by changing the relative angle between lasers, or using different ions (For example the Be$^{+}$ and Yb$^{+}$ ions have very different masses and thus the Lamb-Dicke parameters will be very different if the laser angle remains the same~\cite{JordanBollinger2019,FengMonroe2020,QiaoKim2020}). Here we also note that for very large Lamb-Dicke parameter, both Eq.(\ref{eq:eitnwsc}) and Eq.(\ref{eq:nss}) would no longer be valid. This situation is studied numerically in Ref.~\cite{RoghaniHelm2008}. The ratio $\gamma_g / \gamma_r$ can be tuned by using different ions, for example for Yb$^{+}$ we have $\gamma_g / \gamma_r =1$~\cite{FengMonroe2020}.}
% For the first choice we have $\gamma_{g}/\gamma_{r}=2$ and for the second $\gamma_{g}/\gamma_{r}=1/2$. These two choices would result in the same $\nss$ from Eq.(\ref{eq:eitnwsc}), however from our refined formula in Eq.(\ref{eq:nss}), the first choice would certainly outperform the second one. 
The numerical solutions of the Lindblad master equation are obtained using the QuTip package~\cite{qutip}.

\section{Conclusion}\label{sec:summary}

In summary, we have derived a refined expression for the steady state phonon occupation $\nss$ of the EIT cooling, where the contribution to the second order of the Lamb-Dicke parameter is explicitly given. Based on our analytical result, we show that $\nss$ not only depends on the linewidth $\gamma$ of the excited state, but it also depends on the detailed ratios $\Omega_{d}/\Omega_{b}$ and
$\gamma_{b}/\gamma_{d}$. In particular, we point out that if $\Omega_g \ll \Omega_r$, then $\nss$ diverges when $\gamma_g \rightarrow 0$, thus resolving one difficulty from Eq.(\ref{eq:eitnwsc}) which predicts that $\nss \rightarrow 0$ even in a parameter regime that cooling can not be achieved. We also show that the $\eta^2$ correction term may well be larger than the $\eta$-independent term when $\Omega_g \approx \Omega_r$ or when $\eta_g$ is large, which means that with standard EIT cooling one could already reach the recoil temperature, obscuring the necessity of utilizing more complicated dark-state cooling schemes. \gcc{Our results indicate that to suppress the $\eta^2$ correction term, thus reaching a low steady state phonon occupation in ion cooling experiments, one should properly choose the energy levels as well as the laser strengths such that $\Omega_g \ll \Omega_r$ and that $\gamma_g$ is comparable to or larger than $\gamma_r$.}

% \gcc{This also indicates that to achieve a low steady state phonon occupation in ion cooling experiments, one should properly choose the energy levels such that $\gamma_g$ is at least comparable to $\gamma_r$.} \gcc{On the other hand, from Eq.(\ref{eq:nssa}) one could also supress the contributions from the $\eta^2$ correction term by tuning $\Omega_g \ll \Omega_r$ and $\gamma_g \gg \gamma_r$ at the same time.}

\begin{acknowledgments}
We thank Jie Zhang for fruitful discussion. C. G acknowledges support from National Natural Science Foundation of China under Grants No. 11805279. W. W acknowledges support from National Basic Research Program of China under Grant No. 2016YFA0301903.
\end{acknowledgments}

% We derived the effective Bloch equations, and calculated a more detailed
% steady state occupation than previous result, showing that $\left\langle n\right\rangle _{\textrm{st}}$
% not only depends on the linewidth $\gamma$ and laser detuning $\Delta$,
% but also relies on the ratios $\frac{\Omega_{d}}{\Omega_{b}}$ and
% $\frac{\gamma_{b}}{\gamma_{d}}$. From the analytical results and
% simulations, ion can be cooled to subcoil temperature using EIT cooling.
% by choosing a stronger $\Omega_{g}$ and $\Omega_{g}^{2}\ll\Omega_{r}^{2}$,
% and the energy levels such that, $\gamma_{g}$ should be larger than
% $\gamma_{r}$, one subcoil temperature by EIT cooling. Our result
% is instructive for experimentally implementation.

\bibliographystyle{apsrev4-1}
% \bibliography{refs}


\appendix

\section{Details for deriving the steady state phonon occupation}\label{app:app1}

We denote the Hamiltonian $\LDHop$ and $\Dop$ in the subspace $\mathcal{V}_0$ as $\LDHop^s$ and $\Dop^s$, which can be written as
\begin{align}\label{eq:Hs}
\LDHop^s = & -\Delta\left|e,0\right\rangle \left\langle e,0\right|+\nu\left|d,1\right\rangle \left\langle d,1\right|+\nu\left|b,1\right\rangle \left\langle b,1\right| \nonumber \\
&-\left(\Delta-\nu\right)\left|e,1\right\rangle \left\langle e,1\right|+2\nu\left|d,2\right\rangle \left\langle d,2\right| \nonumber \\
 & +\frac{\Omega_{b}}{2}\left(\left|b,0\right\rangle \left\langle e,0\right|+\left|e,0\right\rangle \left\langle b,0\right|\right) \nonumber \\
 &+\frac{\Omega_{b}}{2}\left(\left|b,1\right\rangle \left\langle e,1\right|+\left|e,1\right\rangle \left\langle b,1\right|\right) \nonumber \\
 & +\im\frac{\eta\Omega_{d}}{2}\left(\left|e,1\right\rangle \left\langle d,0\right|-\left|g,0\right\rangle \left\langle e,1\right|\right) \nonumber \\
 & +\im\frac{\eta\Omega_{d}}{2}\left(\left|e,0\right\rangle \left\langle d,1\right|-\left|d,1\right\rangle \left\langle e,0\right|\right) \nonumber \\
 & +\im\frac{\eta\Omega_{d}}{2}\left(\left|e,1\right\rangle \left\langle d,2\right|-\left|d,2\right\rangle \left\langle e,1\right|\right),
\end{align}
and
\begin{align}\label{eq:Ds}
\Dop^s(\rhoop) = & \frac{\gamma_{d}}{2}\left(2\left|d,0\right\rangle \left\langle e,0\right|\rhoop\left|e,0\right\rangle \left\langle d,0\right|-\{\rhoop, \left|e,0\right\rangle \left\langle e,0\right| \} \right) \nonumber \\
  & +\frac{\gamma_{d}}{2}\left(2\left|d,1\right\rangle \left\langle e,1\right|\rhoop\left|e,1\right\rangle \left\langle d,1\right|- \{\rhoop, \left|e,1\right\rangle \left\langle e,1\right| \}\right) \nonumber \\
  & +\frac{\gamma_{b}}{2}\left(2\left|b,0\right\rangle \left\langle e,0\right|\rhoop\left|e,0\right\rangle \left\langle b,0\right|-\{\rhoop, \left|e,0\right\rangle \left\langle e,0\right|\}\right) \nonumber \\
  & +\frac{\gamma_{b}}{2}\left(2\left|b,1\right\rangle \left\langle e,1\right|\rhoop\left|e,1\right\rangle \left\langle b,1\right|- \{\rhoop, \left|e,1\right\rangle \left\langle e,1\right|\} \right),
\end{align}
respectively. Substituting Eqs.(\ref{eq:Hs}, \ref{eq:Ds}) and all the possible $\Oop$ into the Eq.(\ref{eq:rhoss}), we will get $49$ coupled equations. Here we first consider the $7$ equations by taking $\Oop = \vert j\rangle\langle j\vert$, which are
\begin{align}\label{eq:diags}
&\gamma_{d}\rho_{e0,e0}-\frac{\eta\Omega_{d}}{2}\sigma_{d0,e1}^{x}=0; \\
&\gamma_{b}\rho_{e0,e0}-\frac{\Omega_{b}}{2}\sigma_{b0,e0}^{y}=0;\\
&-\left(\gamma_{d}+\gamma_{b}\right)\rho_{e0,e0}+\frac{\Omega_{b}}{2}\sigma_{b0,e0}^{y}+\frac{\eta\Omega_{d}}{2}\sigma_{e0,d1}^{x}=0; \\
&\gamma_{d}\rho_{e1,e1}-\frac{\eta\Omega_{d}}{2}\sigma_{e0,d1}^{x}=0; \\
&\gamma_{b}\rho_{e1,e1}-\frac{\Omega_{b}}{2}\sigma_{b1,e1}^{y}=0; \\
&-\left(\gamma_{d}+\gamma_{b}\right)\rho_{e1,e1}+\frac{\eta\Omega_{d}}{2}\sigma_{d0,e1}^{x} \nonumber \\
&+ \frac{\Omega_{b}}{2}\sigma_{b1,e1}^{y}+\frac{\eta\Omega_{d}}{2}\sigma_{e1,d2}^{x}=0; \\
&-\frac{\eta\Omega_{d}}{2}\sigma_{e1,d2}^{x}=0,
\end{align}
where we have used $\sigma_{jk}^y = \trace((\im\left|j\right\rangle \left\langle k\right|-\im\left|k\right\rangle \left\langle j\right|)\rhoss)$
and $\sigma_{jk}^x = \trace((\left|j\right\rangle \left\langle k\right|+\left|k\right\rangle \left\langle j\right|)\rhoss)$. Solving the above equations, we get
\begin{subequations}\label{eq:ss1}
\begin{align}
\rho_{e0,e0} &=\rho_{e1,e1}; \\
\sigma_{d0,e1}^{x} &=\sigma_{e0,d1}^{x}; \\
\sigma_{b0,e0}^{y} &=\sigma_{b1,e1}^{y}.
\end{align}
\end{subequations}
Now we substitute Eqs.(\ref{eq:ss1}) into the following $8$ equations
\begin{align}
&\gamma_{b}\rho_{e0,e0}-\frac{\Omega_{b}}{2}\sigma_{b0,e0}^{y}=0; \\
&\gamma_{d}\rho_{e1,e1}-\frac{\eta\Omega_{d}}{2}\sigma_{e0,d1}^{x}=0; \\
&-\frac{\gamma_{d}+\gamma_{b}}{2}\sigma_{b0,e0}^{x}+\Delta\sigma_{b0,e0}^{y}+\frac{\eta\Omega_{d}}{2}\sigma_{b0,d1}^{x}=0; \\
&-\Omega_{b}\rho_{e0,e0}-\Delta\sigma_{b0,e0}^{x}-\frac{\gamma_{b}+\gamma_{d}}{2}\sigma_{b0,e0}^{y} \nonumber \\
&+\frac{\eta\Omega_{d}}{2}\sigma_{b0,d1}^{y}+\Omega_{b}\rho_{b0,b0}=0; \\
&-\frac{\eta\Omega_{d}}{2}\sigma_{b0,e0}^{x}-\nu\sigma_{b0,d1}^{y}+\frac{\Omega_{b}}{2}\sigma_{e0,d1}^{y}=0; \\
&-\frac{\eta\Omega_{d}}{2}\sigma_{b0,e0}^{y}+\nu\sigma_{b0,d1}^{x}-\frac{\Omega_{b}}{2}\sigma_{e0,d1}^{x}=0; \\
&-\eta\Omega_{d}\rho_{e0,e0}+\eta\Omega_{d}\rho_{d1,d1}+\frac{\Omega_{b}}{2}\sigma_{b0,d1}^{y} \nonumber \\
&-\frac{\gamma_{d}+\gamma_{b}}{2}\sigma_{e0,d1}^{x}-\left(\nu+\Delta\right)\sigma_{e0,d1}^{y}=0; \\
&-\frac{\Omega_{b}}{2}\sigma_{b0,d1}^{x}+\left(\nu+\Delta\right)\sigma_{e0,d1}^{x}-\frac{\gamma_{b}+\gamma_{d}}{2}\sigma_{e0,d1}^{y}=0,
\end{align}
together with the conditions $4\Delta\nu=\Omega_{b}^{2}$
and $\eta^{2}\Omega_{d}^{2}\ll\Omega_{b}^{2}$, we get
\begin{align}
\rho_{d1,d1} &\approx \rho_{b0,b0}\left(1+\frac{4\gamma_{d}\nu^{2}\left(\gamma_{d}+\gamma_{b}\right)}{\eta^{2}\Omega_{d}^{2}\Omega_{b}^{2}}\right); \\
\sigma_{d0,e1}^{x} &=\sigma_{e0,d1}^{x} \approx \rho_{b0,b0}\frac{8\nu^{2}\gamma_{d}}{\eta\Omega_{d}\Omega_{b}^{2}}; \\
\sigma_{b1,e1}^{y}&=\sigma_{b0,e0}^{y} \approx \rho_{b0,b0}\frac{8\nu^{2}\gamma_{b}}{\Omega_{b}^{3}}.
\end{align}
Now substituting $\sigma_{d0,e1}^{x},\sigma_{e0,d1}^{x},\sigma_{b1,e1}^{y},\sigma_{b0,e0}^{y}$ back into Eqs.(\ref{eq:rhoss}) in the main text, we get the solutions (Eqs.(\ref{eq:solutions}) in the main text) for the diagonal terms.
% \begin{align}
% \rho_{b0,b0} & = \frac{\eta_{d}^{2}\Omega_{d}^{2}\left(\gamma_{d}+\gamma_{b}\right)}{4\Omega_{b}^{2}\gamma_{d}}\rho_{d0,d0}\\
% \rho_{e0,e0} & = 0\\
% \rho_{d1,d1} & = \frac{\eta_{d}^{2}\Omega_{d}^{2}\left(\gamma_{d}+\gamma_{b}\right)}{4\Omega_{b}^{2}\gamma_{d}}\left(1+\frac{4\gamma_{d}\nu^{2}\left(\gamma_{d}+\gamma_{b}\right)}{\eta_{d}^{2}\Omega_{d}^{2}\Omega_{b}^{2}}\right)\rho_{d0,d0}\\
% \rho_{b1,b1} & = \frac{\eta_{d}^{2}\Omega_{d}^{2}\left(\gamma_{d}+\gamma_{b}\right)}{4\Omega_{b}^{2}\gamma_{d}}\rho_{d0,d0}\\
% \rho_{e1,e1} & = 0\\
% \rho_{d2,d2} & = \frac{\eta_{d}^{2}\Omega_{d}^{2}}{4\Omega_{b}^{2}}\left(\frac{\gamma_{b}}{\gamma_{d}}\right)\rho_{d0,d0}.
% \end{align}
% Using $\rho_{d0,d0} \approx 1$ for the steady state, we obtain the solutions in Eqs.(\ref{eq:solutions}) in the main text.

\end{document}